\documentclass[11pt]{article}
\usepackage{setspace}
\usepackage{fullpage}
\usepackage[left=0.75in,top=1in,right=0.75in,bottom=1in,nohead]{geometry}
\usepackage{mathtools}
\usepackage{amsfonts}
\usepackage{booktabs}
\usepackage{cite}
\usepackage[usenames,dvipsnames]{xcolor}
\newcommand{\Prob}{\mathbb{P}}

\newcommand{\A}{\boldsymbol{A}}

\usepackage{comment}

\def \Prob {\mathbb{P}} 
\def \A {\mathbf{A}}    
\def \P {\mathbf{P}}    
\def \U {\mathbf{U}}    
\def \I {\mathbf{I}} 

\newtheorem{theorem}{\bfseries Theorem}

\begin{document}
\onehalfspacing
\title{Effect of gene-expression bursts on stochastic timing of cellular events}
\author{Khem Raj Ghusinga \\[0.1cm]
{\small Department of Electrical and Computer Engineering} \\
{\small University of Delaware, Newark, DE, USA}\\
{\small \tt \{khem@udel.edu\}}\\[0.5cm]
 Abhyudai Singh\footnote{Corresponding author} \\[0.1cm]
 {\small Department of Electrical and Computer Engineering,} \\
 {\small Department of Mathematical Sciences,} \\
 {\small Department of Biomedical Engineering,} \\
{\small University of Delaware, Newark, DE, USA}\\
{\small \tt \{absingh@udel.edu\}}}
\date{}
\maketitle
\begin{abstract}
Gene expression is inherently a noisy process which manifests as cell-to-cell variability in time evolution of proteins. Consequently, events that  trigger at critical threshold levels of regulatory proteins exhibit stochasticity in their timing. An important contributor to the noise in gene expression is translation bursts which correspond to randomness in number of proteins produced in a single mRNA lifetime. Modeling timing of an event as a first-passage time (FPT) problem, we explore the effect of burst size distribution on event timing. Towards this end, the probability density function of FPT is computed for a gene expression model with burst size drawn from a generic non-negative distribution. Analytical formulas for FPT moments are provided in terms of known vectors and inverse of a matrix. The effect of burst size distribution is investigated by looking at how the feedback regulation strategy that minimizes noise in timing around a given time deviates from the case when burst is deterministic. Interestingly, results show that the feedback strategy for deterministic burst case is quite robust to change in burst size distribution, and deviations from it are confined to about $20\%$ of the optimal value. These findings facilitate an improved understanding of noise regulation in event timing. 
\end{abstract}

\section{Introduction}
Gene expression, the process by which a gene is transcribed to mRNAs and each mRNA is subsequently translated in to proteins, plays a central role in determining cellular behavior. As the biochemical reactions are innately probabilistic, and species such as gene, mRNA, etc. often occur in low copy numbers at single cell level, expression of a gene is a stochastic process \cite{BKC03,RaO05,RaV08,KEB05,SiS13}. Consequently, even if an isogenic population is induced at the same time, two cells might have different evolution of protein level over time.

An important point in triggering of several cellular events is attainment of a threshold level of the protein being expressed. For example, an environmental cue or internal signal usually induces expression of a regulatory protein which subsequently activates an appropriate cellular response. The activation takes place when a certain threshold level of the regulatory protein is achieved \cite{McA97,AKR07}. Other examples of such events include cell-fate decisions \cite{DeW11,ccc04,BSC06,sbs06,lwy15,SGA09,rhb15,kag13,cel04,pih04}, temporal program of gene activation, etc. \cite{NRR07,PeP07}. Because expression of a gene is a stochastic process, a target protein level might be reached at different times in individual cells of an isogenic clonal population induced at the same time.

The timing of such threshold crossing events can be mathematically formulated using a first-passage time (FPT) process. First-passage time is defined as the first time at which a random walker reached a certain critical level, and has been used in several fields to study threshold crossing phenomena  \cite{mvn95,grh97,fat03,ovm10,wed71,lim01,vor10,cak11,chw15,BMN10,IyZ15,dsl15,GhS14}. Particularly, in the context of stochastic gene expression models, previous works have developed exact analytical formulas for FPT moments when gene expresses in translation bursts \cite{SiD14,GhS15,gds16,GhS14}. In these models, one usually assumes a geometrically distributed translation burst distribution which arises from the assumption that both mRNA translation and mRNA degradation are one step processes occurring at exponentially distributed times \cite{ShS08}.

Here, we relax these assumptions and extend the FPT calculations to include any arbitrary burst size distribution. In addition to modeling the mRNA translation and degradation as multistep processes \cite{kag15}, the arbitrary burst size distribution allows one to potentially model post transcriptional regulation as well \cite{ksk15}. We consider a minimal gene expression model that consists of protein arrival in bursts, and its degradation. The arrival rates of bursts is assumed to be dependent on the protein level thereby implementing a feedback transcriptional control. A first-passage problem for this model is formulated, and the probability density function of FPT is determined exactly. Furthermore, an exact analytical formula for the FPT moments is also provided in terms of product of known vectors and inverse of a matrix. These formulas are written as simple series summations for a simple case when the protein of interest is assumed to be stable. Considering different possible distributions of the burst size, we investigate the optimal feedback strategy that might minimize noise in timing of an event around a fixed time. We show that the effect of bursting can deviate the optimal feedback strategy from a no feedback that results from a birth-death model of gene expression. However, these deviations are within $20\%$ of the optimal no feedback strategy.

Remainder of the paper is organized as follows. In section II, we formulate a gene expression model that describes production of a protein in bursts and its degradation. In the next section, FPT computations for this gene expression model are performed. Section IV deals with determining the moments of FPT, particularly the expressions of first two moments for various distributions of burst size. The effect of different burst size distributions on deviations from the optimal feedback strategy are examined in section V. Finally, the results and potential directions of research are discussed in section VI.

\section{Model Description}
We consider expression of a protein from its gene that is induced at $t=0$ as shown in Fig. 1. The model consists of four fundamental components of the gene expression process, namely, transcription (production of mRNAs from gene), translation (production of proteins from a mRNA), mRNA degradation, and protein degradation.  We assume the transcription rate to be an arbitrary function of the protein level which corresponds to feedback regulation. In terms of notations, we denote the transcription rate when protein level $x(t)=i$ by $k_i$, and the protein degradation rate by $\gamma$. Typically the mRNA half-life is much smaller than the protein half-life \cite{Pau05,ShS08,Ber78,Rig79}, and this time-scale separation can be exploited to ignore the mRNA dynamics. As a result the model reduces to a bursty birth-death process in which each transcription event creates a mRNA molecule and it degrades immediately after synthesizing a burst of protein molecules.

\begin{figure}[h]
\centering
\includegraphics[width=0.35\linewidth]{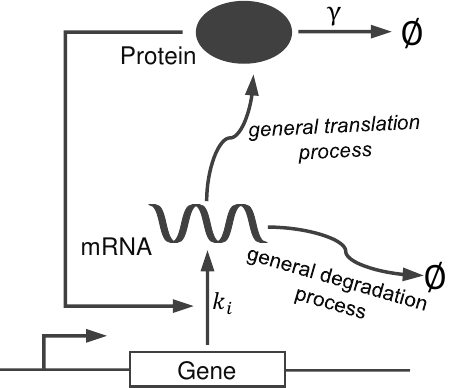}
\caption{Model schematic showing a gene transcribing to mRNAs which are translated into proteins. A feedback regulation is implemented by assuming the transcription rate to be function of the protein level (the transcription rate is $k_i$ when protein level is $i$) . The mRNA translation and degradation processes are assumed to be general, giving rise to some arbitrary non-negative distribution of burst size. Finally, each protein molecule is assumed to degrade with a rate $\gamma$. }
\label{fig:model}
\end{figure}

In essence, the probabilities of occurrence of an arrival event and a degradation event in an infinitesimal time interval $(t,t+dt)$ are given by
\begin{subequations}\label{eqn:bdp}
\begin{align}
\Prob \{x(t+dt)&=i+B | x(t)=i\}=k_idt, \\
\Prob \{x(t+dt)&=i-1 | x(t)=i\}=i\gamma dt,
\end{align}
\end{subequations}
where $B$ represents the burst size. We consider that $B$ follows an arbitrary non-negative distribution as
\begin{subequations} \label{eqn:burstdist}
\begin{align}
\Prob\left(B=i\right) & =\beta_i, \quad i \in \{0, 1, 2, \ldots, \infty \}, \\
\sum_{i=0}^{\infty}\beta_i &= 1.
\end{align}
\end{subequations}
The arbitrary distribution of $B$ allows us to relax assumptions on the translation, and degradation step of mRNA being one step processes \cite{kag15} and also incorporate any post transcriptional regulation \cite{ksk15}. For example, consider one-step degradation of mRNA, i.e., exponentially distributed mRNA lifetime. Denoting the average mRNA lifetime by $1/\tilde{\gamma}$ and the translation rate of a protein from a mRNA by $\tilde{k}$, the burst size distribution can be computed as
\begin{subequations}\label{eqn:burstdistint}
\begin{align}
\Prob\left(B=i\right)&=\int_{0}^{\infty}\tilde{\gamma}e^{-\tilde{\gamma}s}\frac{(\tilde{k}s)^{i}}{i!}e^{-\tilde{k}s}ds \\
			     &=\left(\frac{\tilde{k}}{\tilde{k}+\tilde{\gamma}}\right)^i \left(\frac{\tilde{\gamma}}{\tilde{k}+\tilde{\gamma}}\right)
\end{align}
\end{subequations}
which is a geometric distribution \cite{ShS08}. If instead of an one-step mRNA degradation, a multistep degradation is considered that corresponds to an Erlang distributed mRNA lifetime, then using a similar integral as \eqref{eqn:burstdistint} results in a negative binomial distribution. In the extreme case when the mRNA lifetime is considered to be deterministic, the integral results in a Poisson distributed burst size. 

Another advantage of considering a general burst size distribution is that it also accounts for static extrinsic noise. As an example, we can consider that the burst size follows a geometric distribution given by \eqref{eqn:burstdistint}, and the average burst size is affected by some enzyme $Z$ that is drawn from a distribution, i.e., its levels do not fluctuate over the time-scale of the event. The factor $Z$ here might represent cell-to-cell variability in some factor that affects the translation machinery. The resulting burst size distribution can be written as
\begin{align}
\Prob\left(B=i\right)&=\sum_{z\in \text{support of Z}}\left(\frac{\tilde{k}z}{\tilde{k}z+\tilde{\gamma}}\right)^i \left(\frac{\tilde{\gamma}}{\tilde{k}z+\tilde{\gamma}}\right)\Prob(Z=z),
\end{align}
which can be well described by \eqref{eqn:burstdist}. In the next section,  we compute the first-passage time distribution of the bursty birth-death process. 

\section{First-passage time calculations}
The first-passage time (FPT) is defined as the first-time at which a stochastic process $x(t)$ crosses a threshold $X$. Mathematically, we are interested in determining the probability distribution function (pdf) of the following random variable
\begin{equation}
T=\inf\left\{t\geq 0: x(t) \geq X \right\}.
\end{equation}

As done in our previous work \cite{GhS15}, the first-passage time can be computed by constructing an equivalent bursty birth-death process wherein all states greater than or equal to $X$ are absorbing. The protein count evolves as per probabilities of occurrences given in \eqref{eqn:bdp} until one of the absorbing state is achieved. The difference between this equivalent formulation and the original bursty birth-death process is that in the original formulation, the protein count can return back to the states less than $X$. However, in terms of first-passage times both processes are equivalent. The first-passage time probability density can be computed as
\begin{equation}
f_T(t)=\sum_{i=0}^{X-1}k_i \Prob\left(B\geq X-i\right)p_i(t),
\label{eqn:fptpdf}
\end{equation}
where $p_i(t)=\Prob\left(x(t)=i\right)$. Intuitively, this formula can be interpreted as follows: the process crosses the threshold $X$ for the first time at time $t+dt$ if the protein count was equal to $i$ at time $t$ and a burst of size greater than or equal to $X-i$ occurred in the next infinitesimal time interval $(t,t+dt)$. 

For sake of convenience, we can express \eqref{eqn:fptpdf} as
\begin{subequations}
\begin{equation}
f_\tau(t)=\U^\top \P(t),
\label{eqn:fptpdfvec}
\end{equation}
where 
\begin{equation}
\P(t)=\begin{bmatrix}p_0(t) & p_1(t) & p_2(t) & \ldots & p_{X-1}(t) \end{bmatrix}^\top,
\end{equation}
and $\U$ is given as
\begin{equation} \label{eqn:Ut}
\U=\begin{bmatrix}k_0 \left(1-\sum_{j=0}^{X-1}\beta_j\right)&  \ldots & k_{X-1} \left(1-\beta_0\right)\end{bmatrix},
\end{equation}
where we have used 
\begin{equation}
 \Prob\left(B\geq i\right) =1-\sum_{j=0}^{i-1}\beta_j
\end{equation}
\end{subequations}
which is obtained from \eqref{eqn:burstdist}.

The time evolution of $\P(t)$ can be obtained from the forward Kolmogorov equation (also called the chemical  master equation) for the equivalent bursty birth-death process. This can be written as
\begin{subequations}
\begin{align}
&\dot{p_0}(t) = -k_0(1-\beta_0)p_0(t)+\gamma p_1(t),\\
&\dot{p_i}(t) = -\left(k_i (1-\beta_0) +i\gamma\right)p_i(t)+(i+1)\gamma p_{i+1}(t)  \nonumber \\ 
&\quad \quad \quad \quad +\sum_{n=0}^{i-1} k_n \beta_{i-n}  p_n(t), \quad 1 \leq i \leq X-2, \\
&\dot{p}_{X-1}(t) = -\left(k_{X-1}(1-\beta_0) +(X-1)\gamma\right)p_{X-1}(t)  \nonumber \\  
& \qquad \qquad \qquad +\sum_{n=0}^{X-2} k_n \beta_{X-n-1}  p_n(t).
\end{align}
\end{subequations}
These equations can be compactly written in form of a linear matrix differential equation
\begin{subequations}
\begin{equation}
\dot{\P}=\A \P,
\label{eqn:lindiff}
\end{equation}
where $a_{i,j}$, the element in $i^{th}$ row and $j^{th}$ column of the Hessenberg matrix $A$, is given as
\begin{equation}\label{eqn:matA}
a_{i,j}=
\begin{cases}
0, & j>i+1, \\
\left(i-1\right)\gamma, & j=i+1, \\
-k_{i-1}\left(1-\beta_0\right)-(i-1)\gamma, & j=i, \\
k_{j-1}\beta_{i-1}, & j<i.
\end{cases}
\end{equation}
\end{subequations}
The solution \eqref{eqn:lindiff} is given by the following
\begin{equation}
\P(t)=\exp(\A t)\P(0),
\label{eqn:solLinDiff}
\end{equation}
where $\P(0)$ is the vector consisting of probabilities of the initial protein count. In this work, we consider $x(0)=0$ which leads to 
\begin{equation}
\P(0)=\begin{bmatrix} 1 & 0 & \ldots & 0\end{bmatrix}^\top,
\end{equation}
although any other value of $x(0)$ can be taken as long as it is less than $X$. One can also choose a probability distribution for $x(0)$.

Using \eqref{eqn:solLinDiff} in \eqref{eqn:fptpdfvec} yields the following for the first-passage time probability density function
\begin{equation}
f_T(t)=\U^\top \exp(\A t) \P(0).
\label{pdf}
\end{equation}

\begin{figure}
\centering
\includegraphics[width=0.5\linewidth]{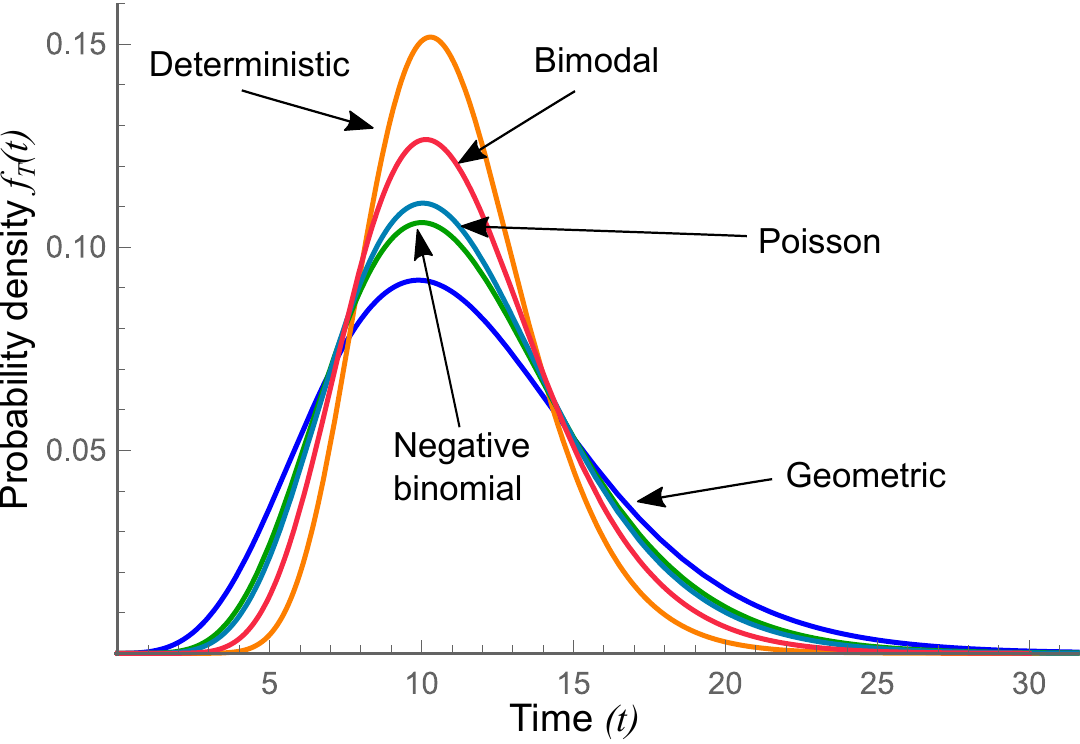}
\caption{First-passage time distributions for different translation burst distributions show qualitatively similar shapes. The mean burst size of $1$ molecule, transcription rate $k_i=2 \text{ min}^{-1}$, degradation rate $\gamma=0.01 \text{ min}^{-1}$ and the event threshold $X=20$ molecules are taken to be same across different distributions. Whereas the Geometric, Poisson, and the Deterministic distributions are uniquely determined by the mean burst size, the Negative binomial and Bimodal distributions have additional parameters. For the Negative binomial distribution the shape parameter is taken to be $5$. For the Bimodal distribution, the probability of having bursts of size $1$ and $2$ are respectively taken to be $0.50$ and $0.25$ whereas probability of having burst of size $0$ is taken to be $0.25$.}
\label{fig:distributions}
\end{figure}

One can numerically compute the pdf in \eqref{pdf} for given model parameters. In Figure~\ref{fig:distributions}, we plot it for several different distributions of the burst size. The production and degradation rates, the event threshold $X$, and the average burst size are taken to be constant across these distributions. It can be seen that the deterministic burst size distribution has a tighter distribution than others -- a feature that is expected. Interestingly, the qualitative shape of the FPT distribution does not appear to vary between different burst size distributions. Next, we discuss how the moments of FPT can be computed using the pdf in \eqref{pdf}.

\section{Moments of first-passage time}
One is often interested in estimating a few low order moments, particularly the average time to an event and its second order moment. Using the probability density function in \eqref{pdf}, the moments of the first-passage time can be calculated as
\begin{subequations}
\begin{align}
\left<T^m\right> &= \int_{0}^{\infty}t^m \U^\top \exp(\A t) \P(0) dt \\
			&= \U^\top \left(\int_{0}^{\infty}t^m \exp(\A t)\right) \P(0) dt.
\end{align}
\end{subequations}
The above integral can be explicitly written in terms of inverse of $\A$
\begin{equation}
\left<T^m\right> =m! (-1)^{m+1} \U^\top \left(\A^{-1}\right)^{m+1}\P(0). \label{eqn:moment}
\end{equation}
as long as the matrix $\A$ is Hurwitz stable. As shown in appendix~\ref{app:hurwitz}, the matrix $\A$ in \eqref{eqn:matA} is indeed Hurwitz stable regardless of the distribution of the burst size. 

To be able to find a certain moment of FPT, one is required to compute $\A^{-1}$. Recall that $\A$ is a lower Hessenberg matrix, i.e., all elements above its first super-diagonal are zero. There are recursion formulas available to invert Hessenberg matrices which can be used to compute \eqref{eqn:moment}, e.g., see \cite{Zho88,Ike79}. In our experience, we have seen that computation of $\A^{-1}$ is well-behaved if one writes
\begin{equation}
\A=\A_0+\A_\gamma,
\end{equation}
and computes $\A^{-1}$ as
\begin{equation}
\A^{-1}=\left(\I+\A_0^{-1}\A_\gamma\right)^{-1}\A_0^{-1}.
\label{eqn:Ainv}
\end{equation}
Here $\A_0$ is the lower triangular matrix which is equal to $\A$ when $\gamma=0$ and $\A_\gamma$ contains the $\gamma$ terms. In some special cases such as when the burst size follows a geometric distribution, an exact expression of $\A^{-1}$ can be obtained and the first two  FPT moments can be further simplified to series summations (see eqs. (16)-(17) in \cite{GhS15}). Generally speaking, the resulting formulas obtained from the above expression are quite convoluted, and it is hard to get any physical insights from them.

It turns out that for a simpler case when $\gamma=0$ (consequently $\A_\gamma=0$), the resulting formulas can be expressed in rather elegant forms as presented below.
\begin{theorem}\label{thm:fptmom}
Consider the FPT probability density function in \eqref{pdf}. When the protein of interest does not degrade, i.e., $\gamma=0$, the first two moments of FPT are given by
\begin{subequations}
\begin{align}
\left<T\right>&=-\sum_{i=0}^{X-1}\frac{\alpha_{i}}{k_{i}} \\
\left<T^2\right>&=2\sum_{i=0}^{X-1}\frac{\alpha_{i}}{k_{i}}\tau_{i}, \quad \tau_{i}=\sum_{j=i}^{X-1}\frac{\alpha_{j}}{k_{j}},
\end{align}
where the coefficients $\alpha_{i}$ depend upon the burst size distribution in \eqref{eqn:burstdistint} and are recursively computed as
\begin{align} 
& \alpha_0=-\frac{1}{1-\beta_0}, \label{eqn:alpha0}\\
& \alpha_i=\frac{\alpha_0 \beta_i + \alpha_1 \beta_{i-1}+\ldots + \alpha_{i-1} \beta_1}{1-\beta_0}, i=1, 2, \ldots, X-1. \label{eqn:alphai}
\end{align}
\end{subequations}
\end{theorem}
\paragraph{Proof:}
When $\gamma=0$, the matrix $\A$ is equal to $\A_0$ and given by
\begin{align}
&\A_0 = \nonumber \\ 
&{\begin{bmatrix}
-k_0(1-\beta_0)                 & 0                                          & \cdots                                              & 0              \\
k_0\beta_1            & -k_1(1-\beta_0)                      & \cdots        & 0                                \\
\vdots                          & \vdots                                            & \vdots         & \vdots                                             \\
k_0\beta_{X-1}  & k_1\beta_{X-2}                     & \cdots                         & -k_{X-1}(1-\beta_0) \\
\end{bmatrix}}.
\end{align}

As $\A_0$ is a lower triangular matrix, its inverse is also lower triangular. It can be shown that $\A_0^{-1}$ has following form
\begin{align}
\A_0^{-1}={\begin{bmatrix}
\frac{\alpha_0}{k_0}                 & 0                                          & \cdots        & 0                                            & 0              \\
\frac{\alpha_1}{k_1}            & \frac{\alpha_0}{k_1}                       & \cdots        & 0                                             & 0              \\
\vdots                          & \vdots                                            & \vdots         & \vdots                                      & \vdots          \\
\frac{\alpha_{X-1}}{k_{X-1}}  & \frac{\alpha_{X-2}}{k_{X-1}}                     & \cdots         & \frac{\alpha_{1}}{k_{X-1}}                & \frac{\alpha_0}{k_{X-1}}  \\
\end{bmatrix}},
\label{eqn:A0inv}
\end{align}
where the coefficients $\alpha_0, \alpha_1, \ldots \alpha_{X-1}$ are determined by the burst size distribution as given in \eqref{eqn:alpha0}-\eqref{eqn:alphai} (also see Remark 1).

Recall from \eqref{eqn:moment} that the mean FPT is given by
\begin{equation}
\left<T\right>=\U^\top \A^{-1} \A^{-1}\P(0).
\end{equation}
Since $\P(0)=\begin{bmatrix}1 & 0 & \ldots & 0\end{bmatrix}^\top$, $\A^{-1}\P(0)$ is just the first column of $\A^{-1}$. Furthermore, as shown in appendix \ref{app:UA0inv} we have that $\U^\top \A^{-1} =-\begin{bmatrix}1 & 1 & \ldots & 1 \end{bmatrix}$. Therefore, $\left<T\right>$ is equal to the negative sum of first column elements of $\A^{-1}$, resulting in
\begin{equation}
\left<T\right>=-\sum_{i=0}^{X-1}\frac{\alpha_{i}}{k_{i}}.
\end{equation}
In the same manner, we can compute the second order moment as
\begin{align}
&\left<T^2\right>=-2\U^\top \A^{-1} \A^{-1} \A^{-1}\P(0) \\
&=2\begin{bmatrix} 1 \\ 1  \\  \vdots \\ 1 \end{bmatrix}^\top
\begin{bmatrix}
\frac{\alpha_0}{k_0}                 & 0                                          & \cdots        & 0                                            & 0              \\
\frac{\alpha_1}{k_1}            & \frac{\alpha_0}{k_1}                       & \cdots        & 0                                             & 0              \\
\vdots                          & \vdots                                            & \vdots         & \vdots                                      & \vdots          \\
\frac{\alpha_{X-1}}{k_{X-1}}  & \frac{\alpha_{X-2}}{k_{X-1}}                     & \cdots         & \frac{\alpha_{1}}{k_{X-1}}                & \frac{\alpha_0}{k_{X-1}}  \\
\end{bmatrix}
\begin{bmatrix} \frac{\alpha_0}{k_0} \\ \frac{\alpha_1}{k_1}  \\  \vdots \\ \frac{\alpha_{X-1}}{k_{X-1}} \end{bmatrix}.
\end{align}
Defining $\tau_{i}=\sum_{j=i}^{X-1}\frac{\alpha_{j}}{k_{j}}$, the above expression simplifies to the following
\begin{equation}
\left<T^2\right>=2\sum_{i=1}^{X}\frac{\alpha_{i}}{k_{i}}\tau_{i}.
\end{equation}

\paragraph{Remark:}The coefficients $\alpha_0,\alpha_1,\ldots$ greatly simplify for some distributions of the burst size $B$. For example, if the burst size is assumed to be drawn from a deterministic distribution such that $\beta_i=1$ when $i=1$ and zero otherwise, then the coefficients $\alpha_{i}=-1$ for all $i=0,1,\ldots, X-1$. On the other hand, if the burst distribution is assumed to follow a geometric distribution with parameter $\mu$ such that $\beta_i=(1-\mu)^i\mu$, then one obtains $\alpha_{0}=-\frac{1}{1-\mu}, \alpha_{i}=-\frac{\mu}{1-\mu}$ for $i\geq 1$. Similarly, for a Poisson distributed burst with parameter $\lambda$, i.e., $\beta_i=\frac{\lambda^{i}}{i!}e^{-\lambda}$, the resulting coefficients are 
\begin{equation*}
\alpha_0=1, \quad \alpha_i=\frac{e^\lambda}{e^\lambda-1}\frac{\lambda^{i}}{i! \left(e^\lambda-1\right)^i}\sum_{m=0}^{i-1}a[i,m]e^{m\lambda},
\end{equation*}
with $a[r,s]$ represents an Eulerian number
\begin{equation*}
a[r,s]=\sum_{i=0}^{s+1}(-1)^i {r+1 \choose i} \left(s+1-i\right)^r.
\end{equation*}

So far we have determined the expressions for FPT moments for a general distribution of the burst size. Using these, we investigate whether for a given burst size distribution there is a specific feedback regulation mechanism that can schedule cellular events with precision. 
\section{Investigating optimal feedback strategy}
As event timing exhibits cell-to-cell variability arising from stochastic nature of gene expression, a problem of interest is to investigate the optimal feedback mechanism that can attenuate variability. Such mechanisms could be employed by cells in cases where precision in timing is important. As an simple example, if one ignores the protein degradation and assumes the burst size to be deterministic, then the FPT moments are given by
\begin{align}
\left<T\right>=\sum_{i=0}^{X-1}\frac{1}{k_{i}}, \quad \left<T^2\right>-\left<T\right>^2=\sum_{i=0}^{X-1}\frac{1}{k_{i}^2}.
\end{align}
For this model, if one minimizes the variance in timing around a fixed mean $\left<T\right>=c$, the optimal transcription rates are given by
\begin{equation}
k_{i}=\frac{X}{c}, \quad i=\{0,1,\ldots,X-1\}.
\end{equation}
Importantly, these transcription rates are equal to each other which represents a no feedback regulation \cite{GFS15}. Furthermore, considering geometrically distributed burst size, and solving for optimal transcription rates yields
\begin{subequations}
\begin{align}
&k_0=\frac{2(2-\mu)+(X-1)\mu}{(2-\mu)(1-\mu)c}, \\
&k_i=\frac{2(2-\mu)+(X-1)\mu}{(1-\mu)c},\quad i=\{1,2,\ldots,X-1\},
\end{align}
\end{subequations}
where $\mu$ is the parameter of geometric distribution \cite{gds16}. Interestingly, here except for the first transcription rate (when the protein level is zero), the other transcription rates are equal, and this optimal feedback strategy is quite similar to a no feedback mechanism. 

Motivated from these findings, we ask how the optimal feedback strategy deviates from a no feedback case for other burst size distributions. To this end, we consider various burst size distributions and numerically find the corresponding optimal feedback strategies. Our results show that even though the optimal feedback strategy shows deviations from a no feedback strategy, these deviations are, however, within 20\% of the transcription rate for no feedback strategy (Fig. 3). Similar to the geometric distributed burst size, the first transcription rate is seen to be significantly different than others even for other distributions. We have also checked the optimal feedback strategies for other values of mean burst size and they are qualitatively similar to the ones shown in Fig. 3.

\begin{figure}
\centering
\includegraphics[width=\linewidth]{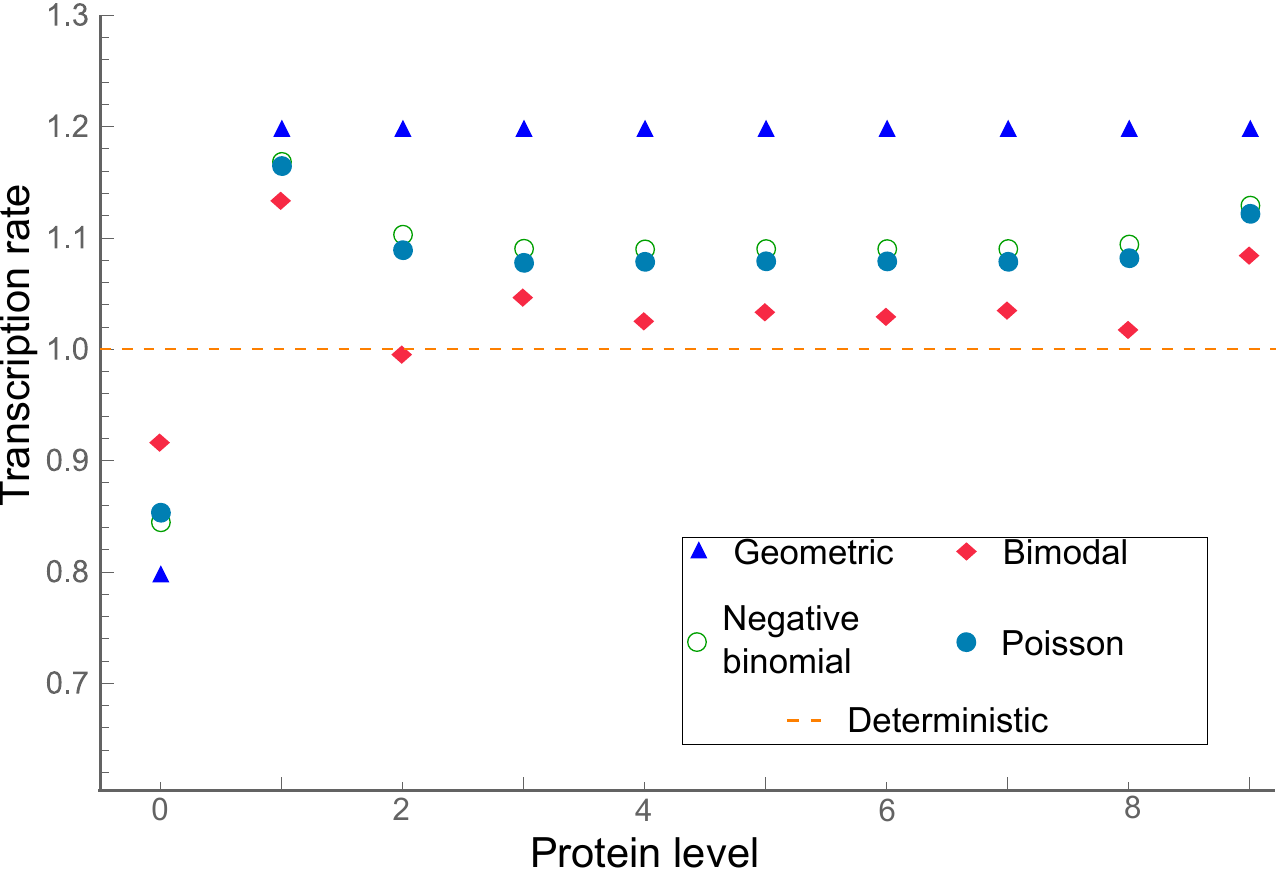}
\caption{Optimal transcription rates for different burst size distributions. The transcription rates are computed by solving an optimization problem that constraints the mean FPT to be $10$ minutes and the event threshold to be $10$ molecules. The mean burst size for each distribution is assumed to be $1$ molecule. While the three distributions (Deterministic, Poisson, and Geometric) are uniquely determined by the mean burst size, the other three require additional parameters. The shape parameter for the Negative binomial distribution is taken to be $5$. For the Bimodal distribution, the probability of having bursts of size $1$ and $2$ are respectively taken to be $0.50$ and $0.25$ whereas probability of having burst of size $0$ is taken to be $0.25$. For the Zipf's law, the exponent is taken to be $10$, and the number of elements is taken to be $30$.}
\label{fig:optzerodeg}
\end{figure}

\section{Conclusion}
Important cellular events are typically governed by accumulation of a regulatory protein up to a critical threshold  \cite{NRR07,PeP07,McA97,AKR07,ccc04,BSC06,sbs06,lwy15,
SGA09,rhb15,kag13,cel04,pih04}. As the expression of the protein is a stochastic process, the resulting timing of events is stochastic as well. Such events can be studied as first-passage time problems. To this end, we considered a stochastic gene expression model that includes translational bursting in protein production and its degradation. One of the typical key modeling assumptions takes the translation and mRNA degradation events as one step process.  The model considered here relaxes this key assumption and considers an arbitrary non-negative discrete distribution of the burst size. We carried out the FPT calculations for this model and found tractable forms of the FPT probability density, and moments. For a special case when the protein of interest does not decay, we found elegant series summations that describe the FPT moments.

Our results show that even though the FPT distribution is affected by the underlying burst size distribution, its shape does not change much with the burst distribution (Fig. 2). Furthermore, for a simple case when protein does not degrade, introduction of burst does quantitatively change the optimal feedback strategy that would minimize the noise in timing around a fixed given time. However, qualitatively, the feedback strategy remains close to a no feedback strategy which is optimal for the case when the burst is deterministic (Fig. 3). These observations suggest the robustness of optimal feedback regulation with respect to burst size distribution.

There are several possible directions of future work. For example, one step could be to take protein degradation into account and perform a systematic analysis of different feedback strategies for various burst size distributions. Furthermore, the production of mRNAs (transcription) and degradation of proteins might also be considered general processes as translation and mRNA degradation. 

\section*{Appendix}
\subsection*{Proof that $\A$ is a Hurwitz matrix}\label{app:hurwitz}
An important requirement for the formula in \eqref{eqn:moment} is that inverse of the matrix $\A$ should exist, and that matrix $\A$ itself must be a Hurwitz matrix.  To show that the matrix is Hurwitz, we prove that it fulfills the following two requirements \cite{LWY07}:
\begin{enumerate}
\item The diagonal elements $a_{ii} < 0$ for $i=1,2,\cdots,X$,
\item $\displaystyle \max_{1\leq j \leq X} \sum_{\substack{i=1 \\j \neq i}}^{X} \left \lvert \frac{a_{ij}}{a_{jj}} \right \rvert < 1$.
\end{enumerate}
As the diagonal elements of $\A$ are $a_{ii}=-k_{i-1}(1-\beta_0)-(i-1)\gamma_{i-1} <0$, the first requirement above is satisfied. To check the second requirement for each of $j=1,2,\cdots,X$, note that
\begin{align}
\sum_{\substack{i=1 \\j \neq i}}^{X} \left \lvert \frac{a_{ij}}{a_{jj}} \right \rvert &= \frac{(j-1)\gamma}{k_{j-1}(1-\beta_0)+(j-1)\gamma} \nonumber+\frac{k_{j-1} \sum_{i=j+1}^{X} \beta_{i-j}}{k_{j-1}(1-\beta_0)+(j-1)\gamma} \\
&\leq \frac{k_{j-1}(1-\beta_0)+(j-1)\gamma}{k_{j-1}(1-\beta_0)+(j-1)\gamma}	=1.	                                                                                             
\end{align}
\normalsize 
Thus $\A$ is a Hurwitz matrix.

\subsection*{Expression of $\U^\top \A_0^{-1}$}\label{app:UA0inv}
Using \eqref{eqn:Ut} and \eqref{eqn:A0inv}, the expression of  $\U^\top \A_0^{-1}$ is
\small{
\begin{align}
&\U^\top \A_0^{-1} = \nonumber \\
&\begin{bmatrix}k_0 \Prob\left(B\geq X\right) \\  \vdots \\ k_{X-1} \Prob\left(B\geq 1\right)\end{bmatrix}^\top
{\begin{bmatrix}
\frac{\alpha_0}{k_0}                 & 0                                          & \cdots        & 0                                            & 0              \\
\frac{\alpha_1}{k_1}            & \frac{\alpha_0}{k_1}                       & \cdots        & 0                                             & 0              \\
\vdots                          & \vdots                                            & \vdots         & \vdots                                      & \vdots          \\
\frac{\alpha_{X-1}}{k_{X-1}}  & \frac{\alpha_{X-2}}{k_{X-1}}                     & \cdots         & \frac{\alpha_{1}}{k_{X-1}}                & \frac{\alpha_0}{k_{X-1}}  \\
\end{bmatrix}},
\end{align}}\normalsize
where $\Prob\left(B\geq i\right)=1-\sum_{j=0}^{i-1}\beta_j$ for $i=1, 2, \ldots, X$. The above expression reduces to
\begin{align}
\U^\top \A_0^{-1} =
\begin{bmatrix}
\sum_{i=0}^{X-1}\alpha_i\left(1-\sum_{j=0}^{X-i-1}\beta_j\right)\\
\vdots \\
\alpha_0\left(1-\beta_0-\beta_1\right)+\alpha_1\left(1-\beta_0\right) \\ \alpha_0\left(1-\beta_0\right)
\end{bmatrix}^\top.
\end{align}
From the relationship between the coefficients $\alpha_i$ and $\beta_i$ in \eqref{eqn:alpha0}-\eqref{eqn:alphai}, it can be established that each of the elements of the above vector is equal to $-1$.

\bibliographystyle{plain}
\bibliography{ACC17Ref}
\end{document}